\documentclass[twocolumn,showpacs,preprintnumbers,amsmath,amssymb, pr]{revtex4-1}

\usepackage{stmaryrd}
\usepackage{txfonts}
\usepackage{amssymb}
\usepackage{mathrsfs}
\usepackage{graphicx}
\usepackage{dcolumn}
\usepackage{bm}
\usepackage{epsfig}
\usepackage{color}                    
\usepackage{hyperref}                 

\begin{document}

\title{Polaron-like vortices, dissociation transition and self induced pinning in magnetic superconductors }

\author{Lev N. Bulaevskii and Shi-Zeng Lin}
\affiliation{Theoretical Division, Los Alamos National Laboratory, Los Alamos, New Mexico 87545}

\begin{abstract}
In magnetic superconductors vortices polarize spins nonuniformly and repolarize them when moving. 
At a low spin relaxation rate and at low bias currents vortices carrying magnetic polarization clouds become polaron-like and their velocities are determined by the effective drag coefficient which is significantly bigger than the Bardeen-Stephen (BS) one. As current increases, vortices release polarization clouds and the velocity as well as the voltage in the I-V characteristics jump to values corresponding to the BS drag coefficient at a critical current $J_c$. The nonuniform components of the magnetic field and magnetization drop as velocity increases resulting in weaker polarization and {\it discontinuous} dynamic dissociation depinning transition. Experimentally the jump shows up as a depinning transition and the corresponding current at the jump is the depinning current. As current decreases, on the way back, vortices are retrapped by polarization clouds at the current $J_r<J_c$. As a result, polaronic effect suppresses dissipation and enhances critical current. Borocarbides (RE)Ni$_2$B$_2$C with a short penetration length and highly polarizable rare earth spins seem to be optimal systems for a detailed study of vortex polaron formation by measuring I-V characteristics. We propose also to use superconductor-magnet multilayer structure to study polaronic mechanism of pinning with the goal to achieve high critical currents. The magnetic layers should have large magnetic susceptibility to enhance the coupling between vortices and magnetization in magnetic layers while the relaxation of the magnetization should be slow. For Nb and proper magnet multilayer structure, we estimate the critical current density $J_c\sim 10^{9}\ \rm{A/m^2}$ at magnetic field $B\approx 1$ T. 
 \end{abstract}
 \pacs{74.25.Wx, 74.70.Dd, 74.25.Ha} 
\date{\today}
\maketitle

\section{Introduction}
The conception of vortex as a polaron \cite{Bulaevskii12a} was initiated by experimental data on the critical current in Er-borocarbide and first we discuss these data. The family of quaternary nickel borocarbides (RE)Ni$_2$B$_2$C (RE is rare earth magnetic ion) is an interesting class of crystals which exhibits both singlet superconductivity and magnetic order at low temperatures. \cite{Canfield98,Budko06,Gupta2006} A number of the crystals in that family develop antiferromagnetic order below the 
N\'{e}el temperature $T_N$, which is below the superconducting critical temperature $T_c$. Because the spatial periodicity of magnetic moments is well below the superconducting correlation length, superconductivity coexists quite peacefully with the antiferromagnetic order. In contrast,  the ferromagnetic order, antagonistic to the Cooper pairing,  leads 
to dramatic changes in both magnetic and superconducting orders in the coexistence phase of singlet superconductors, for a review see Ref. \cite{Bulaevskii85}. 
The compound ErNi$_2$B$_2$C with $T_c=11$ K and $T_N=6$ K attracts numerous attention when it was realized 
that below the phase transition from incommensurate spin density wave (SDW) to commensurate SDW at $T^*=2.3$ K the phase with a weak ferromagnetic ordering may emerge.~\cite{Cho1995,Canfield1996} It was concluded that in ErNi$_2$B$_2$C below $T_N$ the incommensurate SDW develops with effective Ising spins oriented along the $a$-axis and with the wave vector $Q=0.5526~b^*$ from neutron scattering measurements. 
~\cite{Choi2001,Kawano2002} Here $b^*=2\pi/b$ and $b$ is the lattice period along the $b$-axis. At $T^*$ the transition to the commensurate phase with $Q=0.55~ b^*$ leaves one out of 20 spins free of SDW molecular field. These Er spins with the magnetic moment $\mu=7.8\mu_B$ 
are easily polarizable by the magnetic field along the $a$ direction. The spin magnetization in the magnetic field $H=2000$ G in temperature interval 2 K - 4 K follows $M_{{\rm sp}}/H\approx  \mu M_s/(k_BT)$, where $M_s\approx 56$ G, see Fig.~4 in Ref.~\cite{Canfield1996}. This value, $M_s=\mu n$, corresponds to magnetization when all "free" spins with the concentration $n$ order ferromagnetically, The same value $M_s$ was obtained by extrapolation of the magnetization at temperature 2 K in fields $H>1500$ G to $H\rightarrow 0$.  \cite{Gammel2000} Nevertheless, the Hall probe measurements without an applied field below $T^*$ found internal magnetic field much lower than $M_s$ and no spontaneous vortex lattice was seen.~\cite{Bluhm2006} High polarizability of spin system in ErNi$_2$B$_2$C is a key point for our following discussion.

As hope to observe remarkable consequences of weak ferromagnetic phase coexisting with superconductivity waned, a puzzle on ErNi$_2$B$_2$C critical current behavior at low temperatures remained. It was discovered by measuring the hysteresis in the $M-H$ loops and transport measurements that new pinning mechanism develops below 3 K for which the critical current increases as temperature lowers down to $\approx 1.5$ K following approximately the enhancement of magnetic susceptibility.~\cite{Gammel2000,James2001}

To explain these data the conception of vortex as a polaron was proposed, i.e. formation of polaron-like vortices dressed by the polarization cloud
of magnetic moments. \cite{Bulaevskii12a} Generally, the polaronic mechanism is inherent to all magnetic superconductors but  it is most pronounced when the magnetic system is highly polarizable, as in  the case of ErNi$_2$B$_2$C below 2.3 K. To clarify this mechanism, we recall that the magnetic field is nonuniform within the vortex lattice being strongest near the vortex cores. Consequently, the polarization of the magnetic moments is also nonuniform. When vortices move they should repolarize the magnetic system, otherwise they would lose the energy gained by polarization (the Zeeman energy). The 
process of repolarization depends on the dynamics of magnetic system.  In the following we consider the 
relaxation dynamics of free spins in ErNi$_2$B$_2$C. The repolarization process is controlled by the relaxation time $\tau$ which should be compared with the characteristic time 
$a/v$ needed to shift the vortex lattice moving with the velocity $v$ by the vortex lattice period $a=(\Phi_0/B)^{1/2}$. Here $\Phi_0$ is the flux quantum, $B$ is the magnetic induction and we assume a square vortex lattice. For $\tau\gg a/v$ the magnetic moments slow down strongly the vortex motion.
 At some critical velocity and critical current $J_c$, the vortices are stripped off polarization clouds. The corresponding jump in velocity is more evident for a large $\tau$'s. As current decreases, the vortices become retrapped again at the current $J_{r}<J_c$. Since the voltage 
$V\propto v$, the I-V characteristics show hysteresis. The physics here is similar to that of a polaron with vortices playing the role of electrons and the magnetic polarization in place of phonons \cite{AppelBook}.
 
\begin{figure}[t]
\psfig{figure=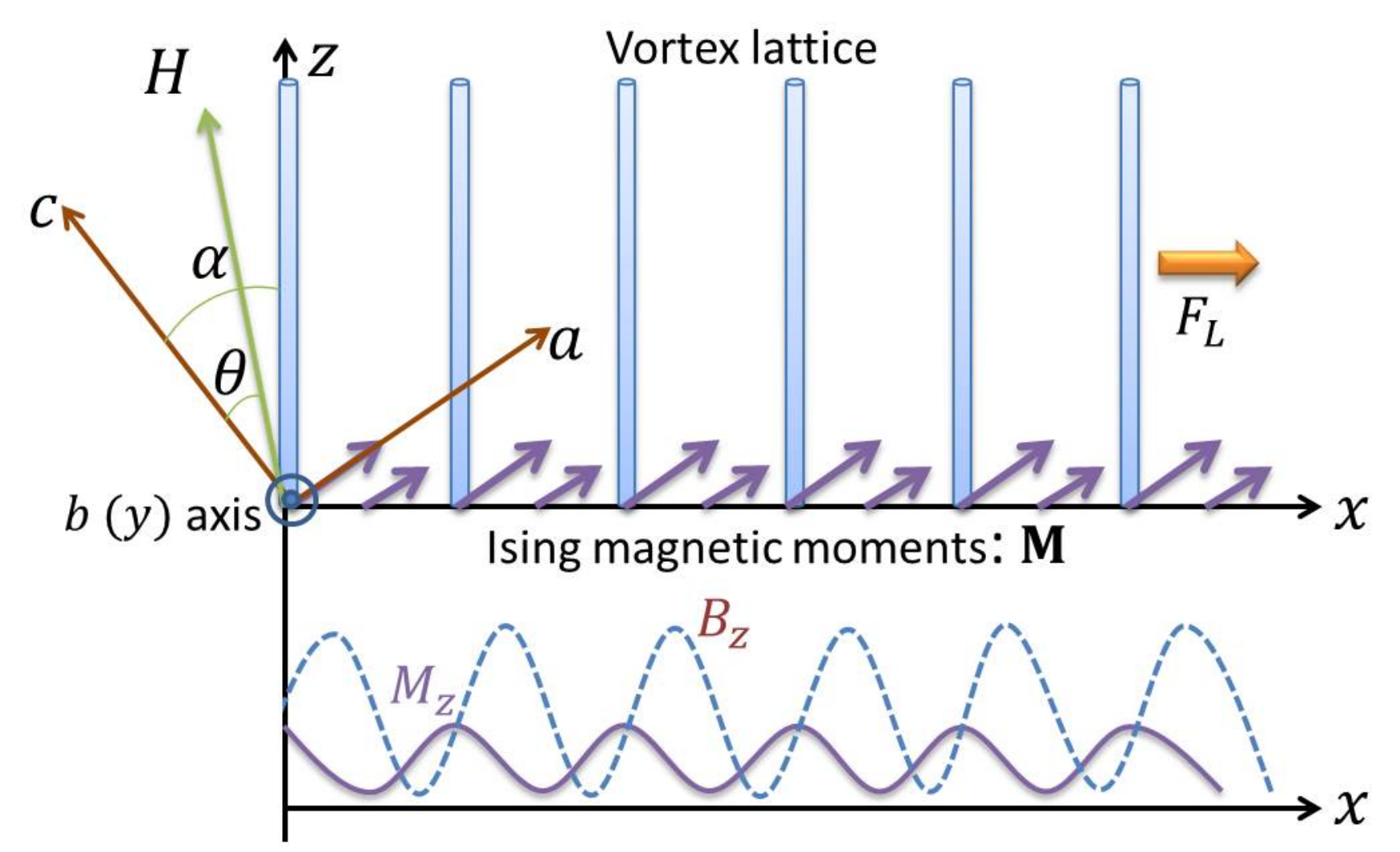,width=\columnwidth}
\caption{\label{f1}(color online) Schematic view of the vortex lattice in the presence of free Ising magnetic moments along the $a$ axis. The vortex lattice is tilted from the applied magnetic fields in the $ac$ plane due to the polarization of the magnetic moments. The vertical columns show the vortex cores. The polarized magnetic moments are nonuniform in space due to the spatial modulation of the vortex lattice magnetic field. Due to the Lorentz force $F_L$ vortices move along the $x$ axis. In moving lattice, there is a phase shift between the magnetic induction (dashed line) associated with the vortex lattice and the magnetization (solid line) caused by the retardation in the response of magnetic moments to the vortex magnetic field.} 
\end{figure}

\section{General equations}

The ErNi$_2$B$_2$C crystals  have orthorhombic structure below $T_N$ with domains where $a$- and $b$-axes change by $90^\circ$ in neighbouring domains. We consider a clean single-domain crystal. In multi-domain crystals, the domain walls also provide a pinning of vortices, which is sharply peaked when the vortex lines are aligned with the domain walls. \cite{Veschunov2007}  We consider the vortex lattice induced by an applied magnetic field ${\mathbf H}$ tilted by the angle $\theta$ with respect to the crystal $c$ axis. As revealed by neutron scattering, vortices form square lattice in ErNi$_2$B$_2$C. \cite{Yaron1996} 
We choose the $z$ axis along the direction of vortex lines at rest and $x$ axis in the $ac$-plane, see Fig. \ref{f1}. The vortex line deviates from the applied field $\mathbf{H}$ due to the magnetic moments \cite{Yaron1996}. The system is assumed to be uniform along the direction of  vortex lines. In static situation the direction of vortex lines is determined by the effective field ${\mathbf H}+4\pi\overline{{\mathbf M}}$. Here $\overline{{\mathbf M}}$ is the spatial average of the magnetization. We denote by $\alpha$ the angle between vortex lines and the $c$ axis.  

The Lagrangian ${\cal L}\{{\bf R}_i(t), M_z({\bf r},t)\}$  for the whole system reads 
\begin{align}\label{eqa1}
\nonumber{\cal L}\{{\bf R}_i(t), M({\bf r},t)\}=
{\cal L}_M\{M_z({\bf r},t)\}+
{\cal L}_v\{{\bf R}_i(t)\}\\+{\cal L}_{{\rm int}}\{M_z({\bf r},t),{\bf R}_i(t)\}+{\cal L}_{vv}\{\mathbf{R}_i\}+\mathcal{L}_F\{\mathbf{J}\},
\end{align}
where ${\cal L}_M\{M_z({\bf r},t)\}$ is the Lagrangian for the magnetic subsystem
\begin{align}\label{eqa2}
{\cal L}_M\{M_z({\bf r},t)\}=-\int d{\bf r}^2M_z^2({\bf r},t)/(2\chi_{zz}),
\end{align}
and ${\cal L}_v\{{\bf R}_i(t), {\bf r}_j\}$ is the Lagrangian for the interaction between vortices and pinning potential due to quenched disorder
\begin{align}\label{eqa3}
{\cal L}_v\{{\bf R}_i(t), {\bf r}_j\}=-\sum_{i,j}U({\bf R}_i-{\bf r}_j).
\end{align}
Here $U({\bf R}_i-\mathbf{r}_j)$ is the pinning potential at ${\bf r}_j$.  Further, ${\cal L}_{vv}\{\mathbf{R}_i\}$ is the vortex-vortex interaction and $\mathcal{L}_F\{\mathbf{J}\}=\sum_i\mathbf{J}\cdot{\mathbf{R}_i}\Phi_0/c$ is the Lagrangian due to Lorentz force in the presence of bias current density $\mathbf{J}$. Here $\chi_{zz}$ is the magnetic susceptibility at the working external magnetic field. It describes response of magnetic moments to nonuniform component of the field induced by vortices.

 In the London approximation the magnetic field of the vortex lattice inside the crystal is $({\bf r}=x,y)$ 
\begin{equation}\label{eqa4}
B_z({\bf r})=\bar{B}_z\sum_{\mathbf G}\frac{\cos({\mathbf G}\cdot{\mathbf r})}{\lambda^2\mathbf{G}^2+1}, 
\end{equation}
where $\bar{B}_z$ is the averaged magnetic induction, ${\mathbf G}$ are reciprocal vectors of the square lattice and $\lambda$ is the superconducting penetration length renormalized by the magnetic moments.
It is given by the expression $\lambda^2=\lambda_L^2(1-4\pi\chi_{zz})$, where $\lambda_L$ describes magnetic field penetration in the absence of the magnetic moments.\cite{Tachiki79,Gray83,Buzdin84,Bulaevskii85,szlin12MFM} 
Note that the magnetic susceptibility $\chi_{zz}=M_z/B_z$ is smaller than $1/(4\pi)$, i.e. $\chi_{zz}<1/(4\pi)$. The magnetic fluctuations $\left\langle M_z M_z\right\rangle\sim \chi_{zz}/(1-4\pi \chi_{zz})$ diverge when $\chi_{zz}\rightarrow 1/(4\pi)$, which indicates instability of the magnetic system \cite{Blount1979}. Here we also ignore anisotropy of the penetration length.

In the Lagrangian the interaction between vortex line at ${\mathbf R}_i=(x_i,y_i)$, and the  magnetic moments is determined by the term 
\begin{equation}\label{eqa5}
{\cal L}_{\rm{int}}\{{\mathbf R}_i,{\mathbf M}_z\}=\int dt\int d{\mathbf r}^2 B_z(\mathbf{R}_i-\mathbf {r},t) M_z({\mathbf r},t),
\end{equation}
where we describe the magnetic moments in the continuous approximation via the magnetization $M_z({\mathbf r},t)$, because distance between free spins, 35 \AA,~\cite{Kawano2002} is much smaller than the London penetration length $\lambda$, about 500 \AA. \cite{Yaron1996}
We ignore the pair breaking effect of the magnetic moments because they suppress Cooper pairing uniformly as distance between free spins is much smaller than the coherence length, and thus the pair breaking effect by the moments does not introduce pinning. 

Both the magnetization and vortices are governed by a relaxation dynamics characterized by the dissipation function ${\cal R}\{{\bf R}_i(t), M_z({\bf r},t)\}={\cal R}_{M_z}+{\cal R}_v$, where  
\begin{align}\label{eqa6}
{\cal R}_{M_z}\{\dot{M}_z({\bf r})\}=\frac{1}{2}\tau \int d{\bf r}^2\dot{M}_z^2({\bf r}), 
\ \ \ \ {\cal R}_v\{\dot{{\bf R}_i}\}=\eta \sum_i\frac{1}{2}\dot{{\bf R}}_i^2.
\end{align}
Here $\tau$ is the relaxation time for a single spin and $\eta$ is the Bardeen-Stephen drag coefficient per unit vortex length,  $\eta={\Phi _0^2}/({2\pi  \xi ^2c^2\rho_n })$ with $\rho_n$ the normal resistivity slightly above $T_c$. The equation of motion for vortices is given by the Euler-Lagrangian equation of motion
\begin{equation}\label{eqa7}
\frac{d}{dt}\frac{\delta {\cal L}}{\delta {\dot{\bf R}}_i}-\frac{\delta {\cal L}}{\delta {\bf R}_i}+\frac{\delta {\cal R}}{\delta \dot{{\bf R}}_i}=0.
\end{equation}
which gives
\begin{align}\label{eqa8}
\nonumber\eta \frac{\partial {\mathbf R}_i}{\partial t}=\frac{\partial{\cal L}_{\rm{vv}}\{{\mathbf R}_i,{\mathbf R}_j\}}{\partial \mathbf{R}_i}+\frac{\partial {\cal L}_{\rm{int}}\{{\mathbf R}_i,{\mathbf M}\}}{\partial {\mathbf R}_i}\\
+\sum_j \frac{\partial U(\mathbf{R}_i-\mathbf{r}_j)}{\partial {\mathbf R}_i }+\mathbf{F}_L,
\end{align}
with $\mathbf{F}_L=\Phi_0 \mathbf{J}/c$ being the Lorentz force.

\begin{figure}[t]
\psfig{figure=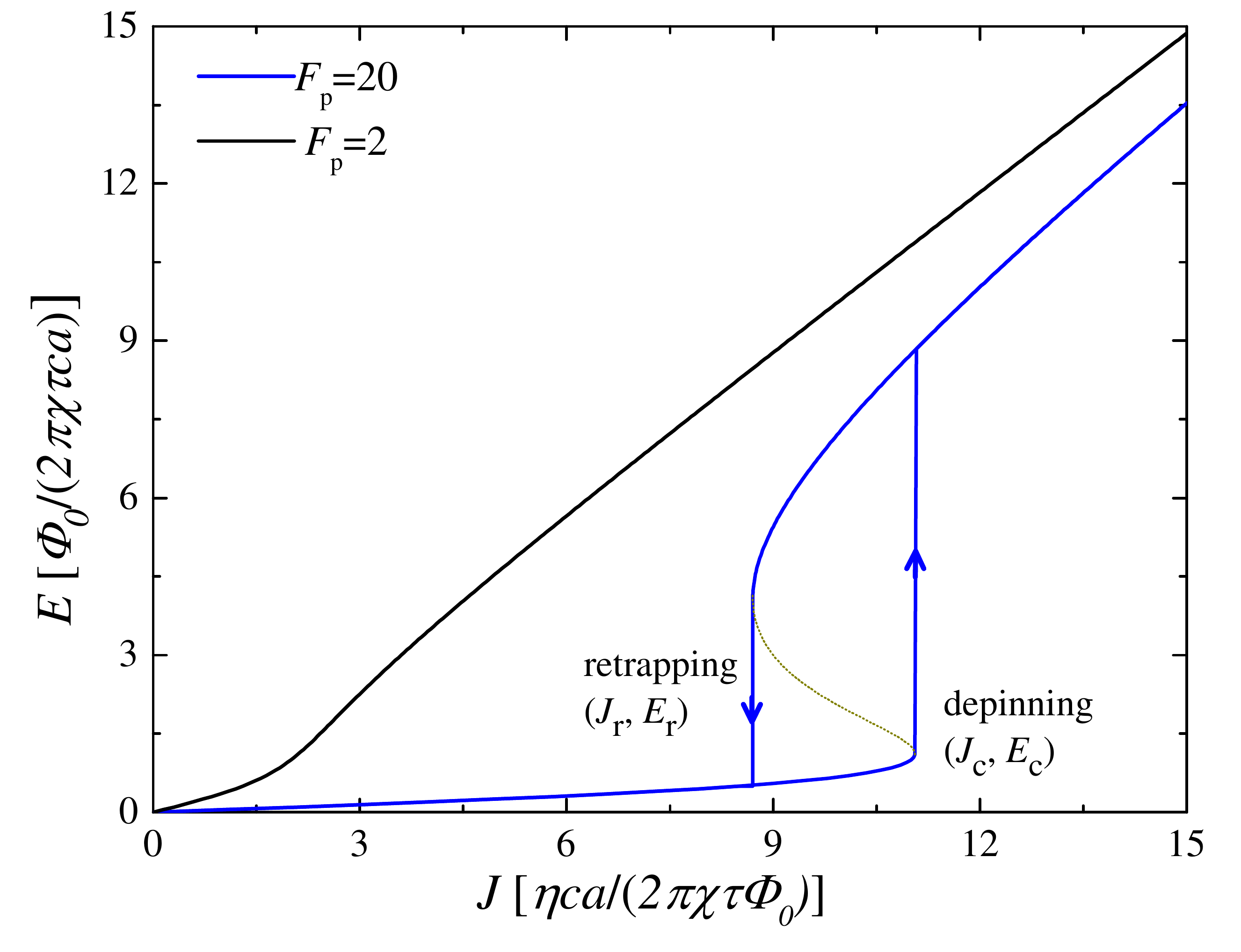,width=\columnwidth}
\caption{\label{f2} Calculated I-V curves for $F_p=20$ and $F_p=2$. For $F_p=20$, the system shows hysteresis in the I-V curve while for $F_p=2$, no hysteresis is present. The dark yellow dotted line denotes the unstable solution.}
\end{figure}
\begin{figure}[b]
\psfig{figure=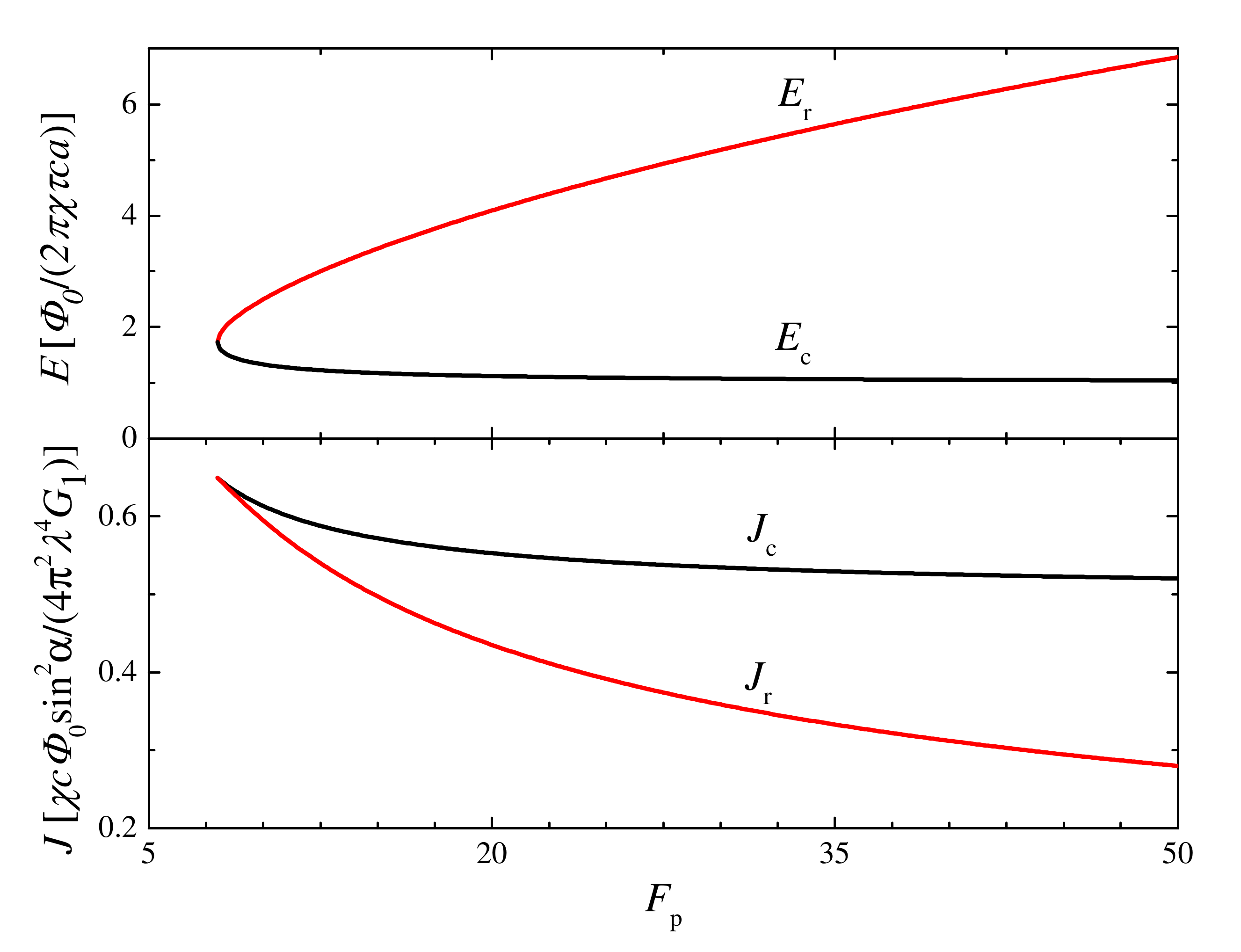,width=\columnwidth}
\caption{\label{f3} Dependence of the critical current $J_c$ and  retrapping current $J_r$, and corresponding electric fields $E_c$ and $E_r$ on $F_p$. When $F_p<8$, hysteresis in I-V curve disappears.}
\end{figure}

We neglect here the effect of quenched disorder because the vortex motion quickly averages out the disorder and the lattice ordering is improved \cite{Koshelev94,Besseling03}. In the lattice phase $\mathcal{L}_{vv}=0$ due to symmetry. The equation of motion for vortex lines then is 
\begin{equation}\label{eqa9}
\eta \frac{\partial {\mathbf R}_i}{\partial t}=\frac{\partial {\cal L}_{\rm{int}}\{{\mathbf R}_i,{\mathbf M}_z\}}{\partial {\mathbf R}_i}+\mathbf{F}_L.
\end{equation}
The dynamics of the magnetization is governed by
\begin{equation}\label{eqa10}
\tau\frac{\partial M_z(\mathbf{r},t)}{\partial t}=-\left[\frac{M_z(\mathbf{r},t)}{\chi_{zz}} -B_z(\mathbf{r})\right].
\end{equation} 
From Eq.~(\ref{eqa10}) we see that relaxation time of the magnetization measured experimentally in the crystal is $\chi_{zz}\tau$. The force due to magnetic moments is the same for all lines and the vortex lattice moves as a whole. The motion of vortex lattice center of mass, $u(t)$, along the $x$-axis is described by the equation 
\begin{equation}\label{eqa11}
\eta \frac{\partial u}{\partial t}=\frac{\partial}{\partial u}\left[\int d{\mathbf r}B_z(x+u,y,t)M_z({\mathbf r},t)\right]+F_L,
\end{equation}
 Using the linear response approach to relate magnetization with the magnetic field, we obtain
\begin{eqnarray}\label{eqa12}
&& \eta \frac{\partial u}{\partial t}=\frac{\partial}{\partial u}\int d{\mathbf r} d{\bf r}'B_z(x+u,y,t)\int_0^tdt'\chi_{zz}({\bf r}-{\bf r}',t-t')\times\nonumber \\
&&B_z({\mathbf r}',t')+F_L,
\end{eqnarray}
The vortex lattice moves with a constant velocity, $u=vt$, in the steady state $t\gg \tau$. Integrating over coordinates and time we obtain
\begin{equation}\label{eqa13}
\eta v=\sum_{{\bf G}}\frac{\chi_{zz}({\bf G}, {\bf v}\cdot{\bf G})}{(\lambda^2{\bf G}^2+1)^{2}}+F_L,
\end{equation}
where $\chi_{zz}({\bf k},\omega)$ is the dynamic magnetic susceptibility in the Fourier representation. We see that the magnetic moments affect strongly vortex motion if a) the resonance Cherenkov radiation condition ${\bf v}\cdot{\bf G}=\Omega({\bf G})$ is fulfilled, where $\Omega({\bf k})$ is the frequency of magnetic excitations with the momentum ${\bf k}$ 
and $\Omega({\bf k})\gg\Gamma({\bf k})$, where $\Gamma({\bf k})$ is the relaxation rate of excitation, 
and b) dynamics of magnetic system is dominated by relaxation, 
$\Omega({\bf k})\lesssim\Gamma({\bf k})$. 
In the former case, discussed in Refs.~\cite{Shekhter11,szlin12a,szln12Spectrum}, the magnetic moments renormalize the vortex viscosity at high velocities when 
alternating magnetic field of vortices is able to excite magnons. Here we consider the latter case of free moments described by the relaxation dynamics according to Eq. \eqref{eqa10} with $\chi_{zz}(\mathbf{k}, \omega)$ given by
\begin{equation}\label{eqa14}
\chi_{zz}(\mathbf{k}, \omega)=\chi\sin^2\alpha\frac{1}{1-i\omega\tau \chi}, \ \ \ \ \chi=\frac{\mu M_s}{k_BT}
\end{equation}
at temperatures $T$ below 3 K for ErNi$_2$B$_2$C.

Renormalizing time in units of $\tau\chi$, length in unit of $1/G_1$, force per unit vortex length in unit of $\eta/(\tau G_1\chi)$ we obtain equation for velocity 
\begin{equation}\label{eqa15}
v+F_p\frac{v}{v^2+1}=F_L,
\end{equation}
where we have accounted only for dominant lattice wave vector ${\bf G}_1=(2\pi/a,0,0)$ and introduced magnetic pinning force per unit vortex length $F_p=\Phi_0^2\tau\chi^2\sin^2\alpha/(4\pi^2\lambda^4\eta)$.
At low bias current (low $F_L$) velocity is proportional to $F_L$ but with enhanced effective viscosity,
$v\approx F_L/(1+F_p)$. At a large $F_L$ (a large $v$) renormalization vanishes and I-V characteristics becomes the usual Bardeen-Stephen one. Importantly, at $F_p>8$ change occurs through sharp transition, as shown in Fig. \ref{f2}. Equation (\ref{eqa15}) at $F_p>8$ for an intermediate $J$ has three real solutions, the highest $v_3$ corresponds to decoupled motion of vortex lattice and magnetization, the lowest $v_1$ corresponds to the motion of vortex-polaron, and the intermediate solution $v_2$ corresponds to an unstable state.

The jump at $J_c$, identifying experimentally as depinning transition, is caused by the dissociation of the vortex-magnon polaron. It is very similar to the dissociation of usual electron-phonon polaron in high electric fields as described theoretically \cite{Banyai1993} and confirmed experimentally in metal oxides \cite{Hed1970}. Upon decreasing the current the vortices are retrapped by the polarization clouds at a threshold current $J_r$ and the vortex lattice moves with a significantly enhanced viscosity at lower currents. The calculated $J_c$, $J_r$ and corresponding electric fields are shown in Fig. \ref{f3}.  At large $F_p$ the critical current is independent of $\tau$:
\begin{equation}\label{eqa16}
J_c\approx 0.03\frac{\chi c\Phi_0\sin^2\alpha}{ G_1 \lambda^4}.  
\end{equation}
$J_c$ decreases with temperature as $J_c\sim 1/T$ and decreases with the magnetic field as $J_c\sim 1/\sqrt{B}$. 

Let us explain the origin of the jumps at $J_c$ and $J_r$. The dependence of the magnetization on the  
velocity of moving vortices is
\begin{equation}\label{eqa17}
{M_z({\bf r},v, t)}={\chi\bar{B}\sin^2\alpha}\sum_{{\bf G}}\frac{\cos[ \mathbf{G}\cdot\mathbf{r}-\beta(v)]}{(\lambda^2{\bf G}^2+1)[1+ (G_1 v \tau\chi)^2]^{3/2}}
\end{equation}
with $\tan(\beta)=G_1 v\tau\chi$. Nonuniform component of the magnetization and thus the polarization effect decrease with velocity. On the other hand, the retardation between the magnetic field and the magnetization, as described by the phase shift $\beta(v)$, increases with the velocity. This positive feedback and the increase of retardation with velocity ensure discontinuous transitions at $J_c$ and $J_r$.

\section{Discussion of experimental data for erbium borocarbide}

A large parameter $F_p$ is required to have a strong pinning due to the polaron mechanism. It is expressed in terms of $\tau$ as $F_p\approx 10^{11}\chi\tau\sin^2\alpha {\rm{s}}^{-1}$, where we have used the coherence length $\xi\approx 13\ \rm{nm}$ \cite{Yaron1996} and the normal resistivity  $\rho_n=5~\mu\Omega\cdot$cm at $T_c$. \cite{Cava1994} The relaxation time $\chi\tau$ in ErNi$_2$B$_2$C is long because the dynamics of majority of spins is strongly suppressed by the formation of the SDW molecular field as was found by the M\"{o}ssbauer measurements.\cite{Bonville1996} The relaxation time drops very fast below 10 K and reaches the value      
$\chi\tau\approx 5\times 10^{-10}$s at $T=5$ K. The data at lower temperatures however were not measured. Thus the only information we have so far is $F_p>50\sin^2\alpha$. 

The critical current for ErNi$_2$B$_2$C reported in Ref.~\cite{Gammel2000} is about 250 A/cm$^2$ for $B=0.1\ \rm{T}$, $T=2\ \rm{K}$. which corresponds to $\alpha=2.5^{\circ}$ according to Eq. ~(\ref{eqa16}). The applied magnetic field was close to the $c$ axis in experiment, but the precise angle $\theta$ was not reported.~\cite{Gammel2000}. The estimate of the order 1$^{\circ}$ is reasonable, but the quantitative comparison is not convincing as we do not know $\tau$ and thus $F_p$ below 2.3 K. We predict hysteretic behavior in ErNi$_2$B$_2$C, strong dependence of voltage and of the critical current on the angle $\theta$, at least for $\theta\gg 0.15^{\circ}$. Hence, the real check of polaronic mechanism should be by measuring the I-V characteristics. We estimate the critical current reaches values as high as 10$^6$ A/cm$^2$ at high angles at $T=1$ K and $B=0.1$ T. 

The effect of ordered spins on the vortex motion is
similar to that described in Refs. \onlinecite{Shekhter11,szlin12a,szln12Spectrum} for an antiferromagnet. When the Cherenkov radiation condition 
${\bf v}\cdot{\bf G}=\Omega({\bf G})$ is satisfied, excitation of magnons results in enhanced drag coefficient by transferring energy from vortex motion to the magnetic subsystem. This occurs at high velocities (high currents) of vortices, due to a gap in the magnon spectrum and a large velocity of magnon, leading to a voltage drop in comparison with the bare BS result.

In the incommensurate SDW when $T>T^*$, some spins experience quite weak SDW molecular field. Thus, they are polarized by vortices and exhibit polaronic effect and pinning. This accounts for the increase of pinning in ErNi$_2$B$_2$C as $T$ decreases below $T_N$, see Ref.~\cite{Gammel2000}, and also the pinning in the holmium borocarbide below $T_N$. ~\cite{Dewhurst1999}

Next we discuss the effect of quenched disorder. In the presence of quenched disorder, the vortex lines adjust themselves to take the advantage of the pinning potential, which destroys the long-range lattice order. Below a threshold current, vortices remain pinned (actually they creep between pinning centers due to fluctuations), where the polaronic mechanism does not play a role in this region. When the current is high enough to depin the vortices from quenched disorder, vortices start to move and the lattice ordering is improved. The vortex viscosity is enhanced by formation of polaron with the nonuniformly induced magnetization. The polaron dissociates and the system jumps to the conventional BS branch at a critical velocity (current). Pinning due to quenched disorder works in the static region and polaronic pinning works in the dynamic region. The critical current of the whole system therefore is the sum of these two threshold currents. Note that magnetostriction in combination with quenched disorder enhance the polaronic pinning mechanism.

\section{Response of the Vortex lattice to an ac driving current}

Here we study the response of a vortex-polaron to an ac driving current \cite{Bulaevskii12PRB}. We write the equations of motion for magnetization $m(t)=M(G_1,t)\lambda^2G_1^2/\Phi_0\chi$ and the vortex lattice center of mass $u(t))$ as
\begin{equation}\label{eqb1}
\partial _t m(t)=-[m(t)-\exp [-i u(t)]],
\end{equation}
\begin{equation}\label{eqb2}
\partial _t u=F_L-\text{Im}\left[ F_p\exp (i u)m(t)\right],
\end{equation}
with an ac Lorentz force $F_L=F_{\text{ac}}\sin (\omega  t)$. Eliminating $m(t)$ we obtain equation for $u(t)$:
\begin{equation}\label{eqb3}
\frac{du}{dt}=F_L-F_p\int _0^tdt'\sin[u(t)-u(t')]\exp(t'-t).
\end{equation}

\begin{figure}[t]
\psfig{figure=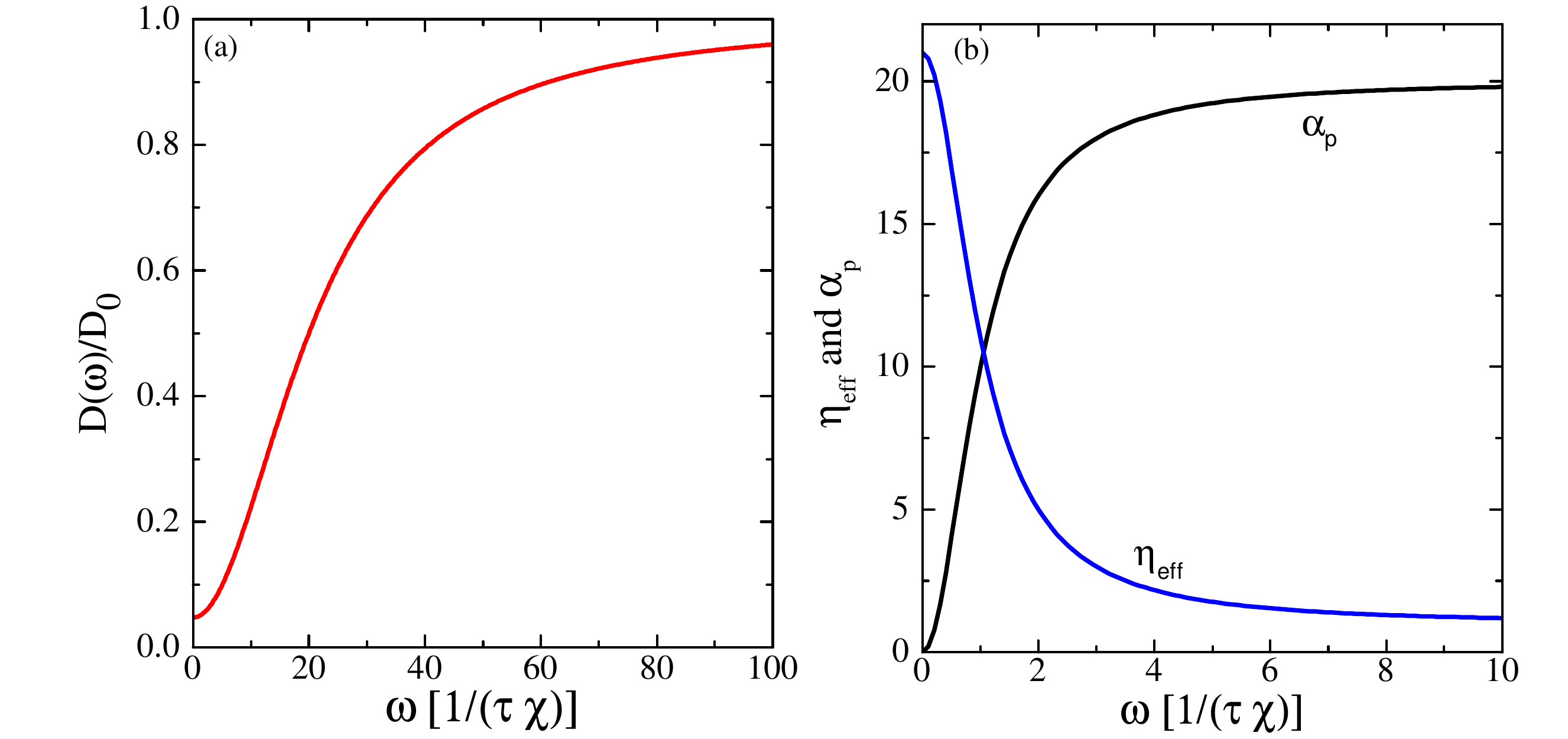,width=\columnwidth}
\caption{\label{f4}(color online) (a) Dependence of the normalized dissipation power, $D(\omega)/D_0$, and (b) the effective viscosity $\eta_{{\rm eff}}$, pinning strength $\alpha_p$ on the driving frequency $\omega$ in the linear regime $F_{ac}\ll \omega(1+F_p)$. Here $F_p=20$. }
\end{figure}

First we consider an ac current regime with a low amplitude $F_{ac}/[\omega(1+F_p)]\ll 1$. Then the vortex lattice oscillates, $u={\rm{Re}}[u_{ac}\exp(i \omega t)]$, with the amplitude
\begin{eqnarray}\label{eqb4}
&&u_{ac}=F_{\text{ac}}(i \eta_{{\rm eff}}\omega +\alpha_p)^{-1}, \\
&&\eta_{{\rm eff}}=1+F_p(\omega^2+1)^{-1}, \ \ \ \ \ \alpha_p=F_p\omega^2(\omega^2+1)^{-1}.
\end{eqnarray}
For a high frequency $\omega\gg 1$, the effect of magnetization is to introduce the pinning potential $U_M=F_pu^2/2$ with strength $F_p$. In this case, the vortex lattice follows the driving force much faster than magnetization, which remains almost time independent. The polarization of the magnetization results in periodic pinning potential with the periodicity of  vortex lattice as it was induced by the same lattice at previous positions and previous moments of time.  For a low frequency $\omega\ll 1$, the effect of magnetization is to renormalize the drag coefficient from $\eta$ to $\eta_{\rm{eff}}=1+F_p$. In this polaron region, the magnetization follows vortex motion by formation of vortex polaron, as in the dc case $\omega=0$, resulting in enhancement of viscosity and suppression of ac dissipation.
The dissipation power of the whole system, averaged over time, $D(\omega)=\langle F_L(t)v(t)\rangle_t$, is reduced due to the presence of magnetic moments. In the linear region with vortex polaron, we obtain 
\begin{equation}\label{eqb5}
D(\omega)=\frac{F_{{\rm ac}}}{2} \frac{\omega^2\eta_{{\rm eff}}}{\alpha_p^2+\eta_{{\rm eff}}^2\omega^2}.
\end{equation}
This dissipation power should be compared with the case without magnetic moments (at $F_p=0$), $D_0=F_{{\rm ac}}^2/2$.  For $\omega\gg 1$
we get $D/D_0\approx 1$ and for $\omega\ll1$ we get $D/D_0=(1+F_P)^{-1}$. The frequency dependence of the normalized dissipation power $D(\omega)/D_0$, effective viscosity $\eta_{{\rm eff}}$, and pinning strength $\alpha_p$ is shown in Fig.~\ref{f4}. The dissipation of the system in the presence of magnetic subsystem is strongly reduced in linear regime $F_{{\rm ac}}<F_{Lc}$, which might be useful for applications.

 \begin{figure}[b]
\psfig{figure=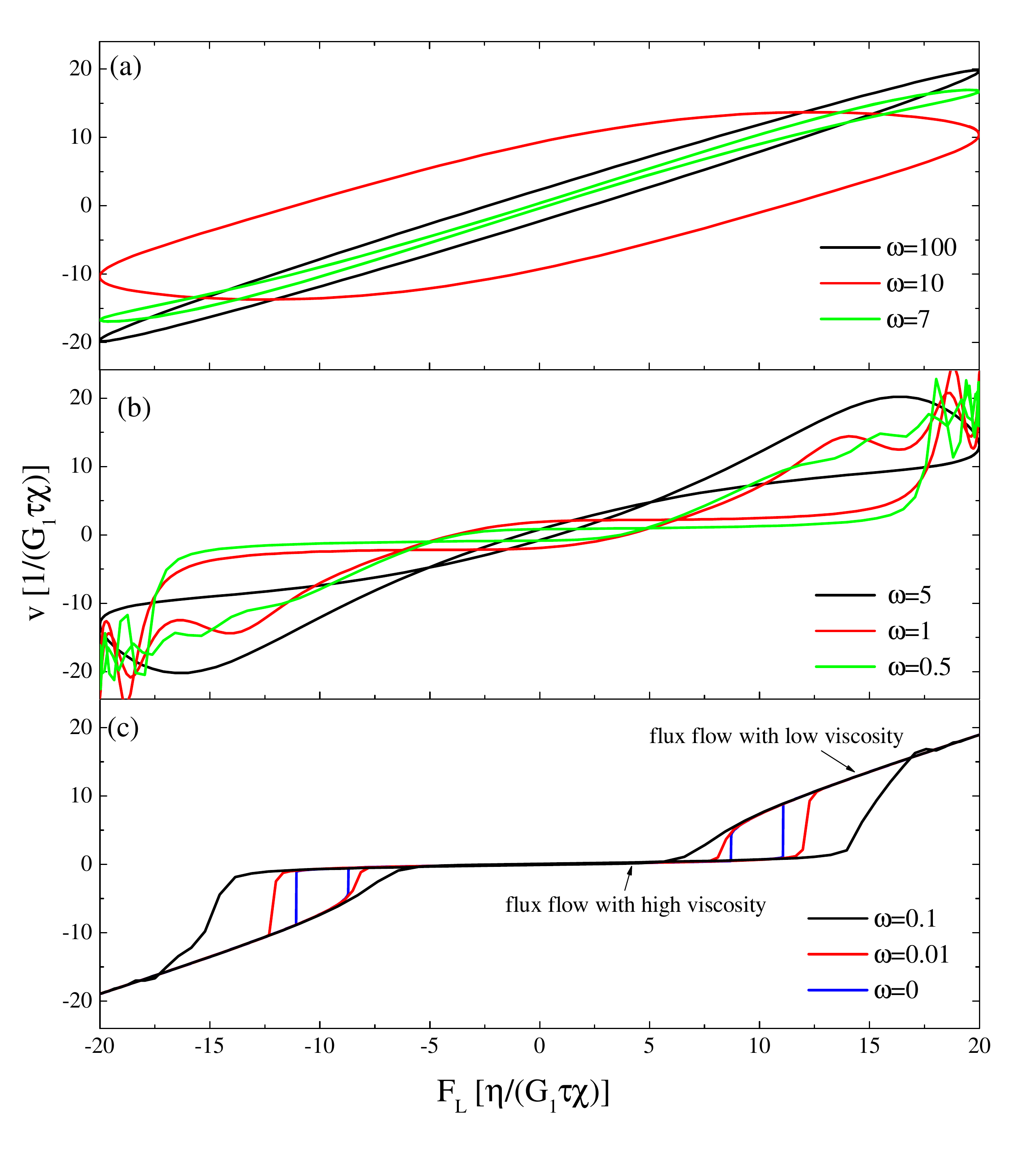,width=\columnwidth}
\caption{\label{f5}(color online) Dependence of the vortex velocity $v(t)$ on the driving force $F_L(t)=F_{{\rm ac}}\sin(\omega t)$ with $F_{{\rm ac}}=20$. Here $F_p=20$.}
\end{figure}

Next we consider stronger driven force amplitude.  In this hysteretic regime, we describe the system analytically  in the adiabatic limit, $\omega\ll 1$. At the time 
moment $t_c$, when $F_L(t)=F_{Lc}\approx  0.5 F_p$, polaron dissociation leaves magnetization and vortex lattice weakly coupled because lattice moves now with a high velocity. The magnetization  component 
$m(t)$ after that moment  relaxes as $m(t)=\exp(-t+t_c)$, and motion of vortex lattice is determined by the equation
\begin{equation}\label{eqb6}
\frac{du}{dt}=F_{Lc}+F_p\sin ( u)\exp(-t+t_c).
\end{equation}
When $t-t_c< 1$ the velocity of the vortex lattice oscillates with the frequency $\Omega=F_{Lc}$, 
\begin{equation}\label{eqb7}
v=F_{Lc}+F_p\sin(F_{Lc} t)\exp(-t+t_c),
\end{equation}
but oscillations relax on the time scale of unity. 
These post dissociation oscillations are caused by the motion of vortex lattice in periodic potential induced by remnant retarded magnetization. 

\begin{figure}[t]
\psfig{figure=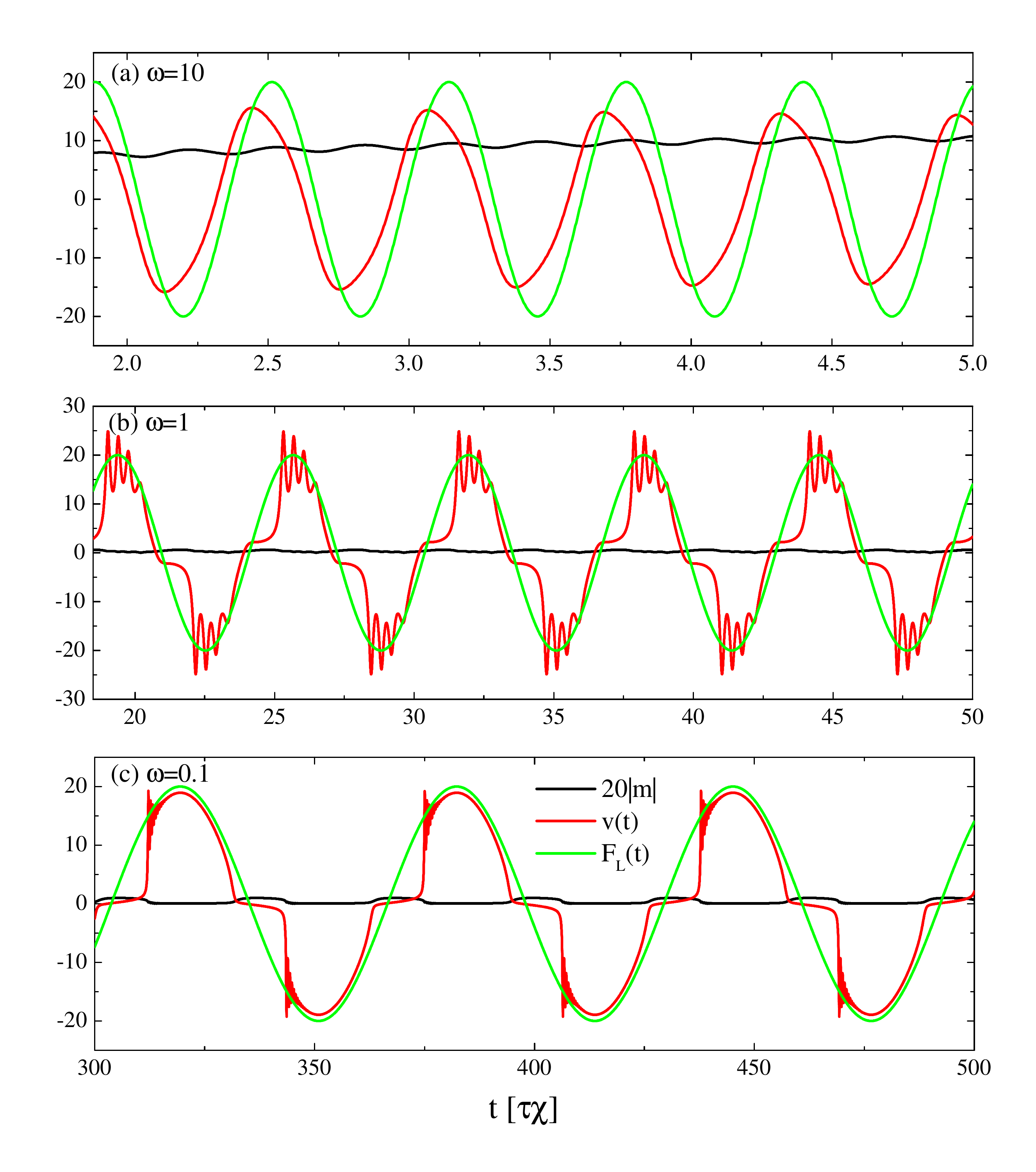,width=\columnwidth}
\caption{\label{f6}(color online) Time evolution of the vortex velocity, $v(t)$, and of magnetization $|m(t)|$ in the presence of ac driving force $F_L(t)=F_{{\rm ac}}\sin(\omega t)$ at several frequencies $\omega$. (a): $\omega=10$, (b): $\omega=1$, (c): $\omega=0.1$. We take $F_{{\rm ac}}=20$ and $F_p=20$.}
\end{figure}

To take both the retardation and nonlinearity into account for an arbitrary $\omega$
we solve numerically Eqs. (\ref{eqb1}) and (\ref{eqb2}).  We consider the interesting region $F_p> 8$, where the dissociation of vortex polaron is possible due to nonlinear effects at $u\geq 1$. We take $F_p=20$ in the following discussion. The hysteretic behavior of vortex lattice velocity vs. driving force is shown in Fig.~\ref{f5}. At frequencies $\omega\lesssim 1$ which are similar to the dc case $\omega=0$, we see the following sequence of events during the period of $F_L(t)$: polaron formation near low $|F_L|$ (interval of low vortex velocity), then polaron dissociation (velocity sharp increase) followed by region of vortex oscillations on the background of average high velocity, decrease of velocity as the Lorentz force drops and vortex retrapping (sharp drop in vortex velocity) and then dissociation again at negative $-F_{Lc}$ (sharp drop in velocity). The results for 
behavior of vortex velocity in time, $v(t)$, at $F_{\rm{ac}}=20>F_{Lc}$ and different $\omega$ are shown in Fig. ~\ref{f6}. 

At all frequencies $\omega\lesssim 1$ we see post dissociation oscillations caused by motion of decoupled  vortices with respect to periodic potential created by nonuniform magnetization  
induced by the same lattice just before decoupling (when velocity was still low) and frozen for some periods of time after decoupling due to the retardation effect. This self-induced pinning due to the retardation, and the amplitude of corresponding vortex oscillations reach maximum at $\omega\approx 1$. In a rough approximation we describe them by the equation
\begin{equation}\label{eqb8}
\frac{du}{dt}\approx F_{Lc}+F_p m_d\sin(u-u_d),
\end{equation} 
assuming approximately constant $m$ and $F_L=F_{Lc}$ in the regions of maxima and minima of the Lorentz force. Here $u_d$ and $m_d$ are the position of vortex lattice and the amplitude of magnetization at the moment of decoupling.
This gives approximate solution:
\begin{equation}\label{eqb9}
v(t)\approx F_{Lc}+F_pm_d\sin(F_{Lc}t),
\end{equation}
which provides rough estimate for the oscillation frequency, $\Omega\approx F_{Lc}$,  when number of oscillations of velocity per the half period of $F_L(t)$ is significantly bigger than unity. This expression for the frequency in original units reads $\Omega\approx 2\pi F_{{\rm ac}}/a\eta$. Such relation is anticipated for decoupled vortex moving in the pinning potential with periodicity $a$.

\section{Enhancement of critical current density in superconducting/magnetic multi-layers}

\begin{figure}[b]
\psfig{figure=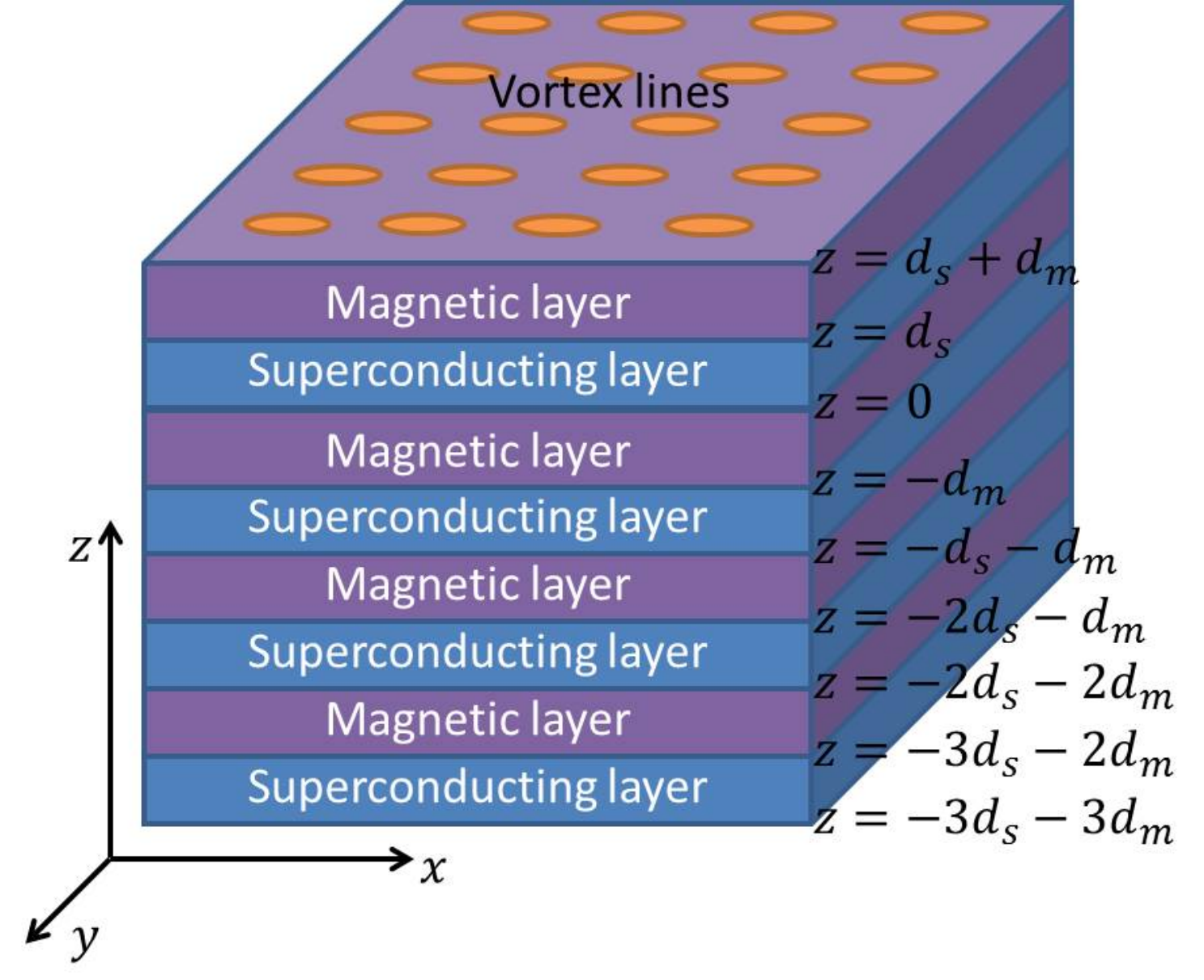,width=\columnwidth}
\caption{\label{f7}(color online) Schematic view of multi-layer structure consisting of alternating magnetic (M) layers (purple) with thickness $d_m$ and superconducting (S) layers (blue) with thickness $d_s$. } 
\end{figure}

The polaronic mechanism of pinning is promising for achieving a high critical current. We propose to use a  superconducting (S) and magnetic (M) multilayer structure as shown in Fig. \ref{f7} to optimize such pinning mechanism. \cite{szln12ML} To achieve high critical current the magnetic layers should have a slow relaxation of the magnetization. Secondly, the magnetic layers should have a high magnetic susceptibility at working magnetic field to ensure a strong coupling between magnetic moments and vortices.  Thirdly, the penetration depth of the superconducting layers should be small, such that the magnetization polarization varies rapidly in space.

The vortex lattice is induced inside the S layers under external magnetic fields. The vortex lattice moves in response to the Lorentz force when a transport current is present. In the quasistatic approximation, the motion of vortex lattice is given by
\begin{equation}\label{eqc1}
\lambda^2\nabla \times \nabla \times \mathbf{B}+\mathbf{B}=\Phi _0\sum_i\delta \left[\mathbf{r}-\mathbf{r}_i(t)\right]\hat{\mathbf{z}},
\end{equation}
where $\hat{\mathbf{z}}$ is the unit vector along the $z$ axis and $\mathbf{r}_i(t)=\mathbf{r}_0-\mathbf{v} t$ is the vortex $i$ coordinate. The magnetic field inside the M layers is governed by the Maxwell equations
\begin{equation}
\nabla\times (\mathbf{B}-4\pi\mathbf{M})=0, \ \ \nabla\cdot \mathbf{B}=0.
\end{equation}
The dependence of magnetization $\mathbf{M}$ on $\mathbf{B}$ is determined by the material properties.  With a strong field and in static case, $\mathbf{M}$ is a nonlinear function of $\mathbf{B}$ and generally can be expressed as $\mathbf{M}(\mathbf{r})=\int d\mathbf{r}^3 f[\mathbf{r}-\mathbf{r}', \mathbf{B(r')}]$.  The characteristic length of magnetic subsystem is much smaller than $\lambda$ and we use a local approximation $f[\mathbf{r}-\mathbf{r}', \mathbf{B(r')}]=\delta(\mathbf{r}-\mathbf{r}')f(\mathbf{B(r')}$. Induction $\mathbf{B}(\mathbf{r})$ has component uniform in space, ${\mathbf{B}_0}$, and the  nonuniform component nonuniform, $\tilde{\mathbf{B}}(\mathbf{r})\ll {\mathbf{B}_0}$. Thus the spatially nonuniform magnetization $\tilde{\mathbf{M}}(\mathbf{r})$ is $\tilde{\mathbf{M}}(\mathbf{r})\approx [\partial f(\mathbf{B}_0)/\partial B_0 ]\tilde{\mathbf{B}}(\mathbf{r})\equiv \chi_0 (\mathbf{B}_0) \tilde{\mathbf{B}}(\mathbf{r})$. In the following we consider isotropic magnetic subsystem characterized by a susceptibility $\chi_0(\mathbf{B}_0)$ at $\mathbf{B}_0$ in static case. The magnetic field inside the M layer is determined by the equation $\nabla^2\tilde{\mathbf{B}}=0$. Since only the spatially nonuniform component $\tilde{\mathbf{M}}$ and $\tilde{\mathbf{B}}$ are responsible for pinning, we will focus on the nonuniform components in the following calculations and omit mark tilde. At the interface between the M and S layers, we use the standard boundary condition for the field parallel to the $z$-axis $B^z$ and field parallel to the interface $B^{||}$ 
\begin{equation}\label{eqc2}
B^z|_{\rm{S}}=B^z|_{\rm{M}},\ \ \ \ B^{||}|_{\rm{S}}=(1-4\pi\chi_0)B^{||}|_{\rm{M}}.
\end{equation}
Then we obtain the magnetic field inside the M layers
\begin{equation}\label{eqc3}
B_m^z(G>0, z)=\alpha\left[e^{G z'}+e^{- G(z'+d_m)}\right]\frac{\Phi _0\exp \left(-i G_x v_x t\right)}{1+\lambda ^2G^2},
\end{equation}
\begin{equation}\label{eqc4}
B_m^{||}(G>0, z)=i \alpha\left[e^{G z'}-e^{- G(z'+d_m)}\right]\frac{\Phi _0\exp \left(-i G_x v_x t\right)}{1+\lambda ^2G^2},
\end{equation}
\[
\alpha=-\frac{ e^{ d_m G} \left(-1+e^{d_s k_s}\right) \chi '}{(1-\chi')(e^{d_s k_s}-e^{G d_m})+(1+\chi')(1-e^{d_m G+d_s k_s})},
\]
with $z'=z-n(d_s+d_m)$, $\ k_s=\sqrt{\lambda^{-2}+G^2}$, and $\chi'=(1-4\pi\chi_0)^{-1}k_s/G$.
Here $n$ is the layer index and the vortex motion is assumed to be along the $x$ direction. We consider square lattice $\mathbf{G}=(m_x2\pi/a, m_y 2\pi/a)$ with $a=\sqrt{\Phi_0/B_0}$ the lattice constant and $m_x$, $m_y$ integers. 

We assume a relaxational dynamics for the M layers, $\mathbf{M}(\omega)=\chi(\omega) \mathbf{B}_m(\omega)$,  with a dynamic susceptibility governed by a single relaxation time $\chi_0\tau$ as in Eq. \eqref{eqa10}. In the steady state, we have 
\begin{equation}\label{eqc6}
\mathbf{M}\left(\mathbf{G}, z, t\right)=\tau^{-1}\int _0^t\exp [(t'-t)/(\chi_0\tau) ]\mathbf{B}_m(\mathbf{G}, z, t')dt'.
\end{equation}
$\mathbf{M}$ depends on the history of vortex motion for a slow relaxation of magnetization, and there is retardation between the time variation of induced nonuniform magnetization and vortex motion. As a result, the magnetization exerts a drag force to the vortex which is opposite to the driving force. The pinning force acting on a single vortex due to the induced magnetization in one M layer is given by $F_M=\partial_{r_0}\int dxdy\int_{-d_m}^0 dz\mathbf{M}\cdot\mathbf{B}_m$, which yields
\begin{equation}\label{eqc7}
F_M=\sum_G \left[1-\exp \left(-2G d_m\right)\right]\frac{2\alpha^2\chi _0}{\left(1+\lambda ^2G^2\right)^2 a^2}\frac{ G v \chi _0\tau \Phi _0^2}{1+ (G v \chi _0\tau)^2 }.
\end{equation}
The I-V curve is determined by the equation of motion for vortex $d_s\eta v=d_s F_L -F_M$ with the electric field $E=B v/c$ and the Lorentz force $F_L=J\Phi_0/c$. We consider a realistic case where $a/(2\pi)\ll d_m, d_s$. Taking into account only the dominant contribution $G_x=2\pi/a$ and $G_y=0$ in the summation, we obtain the same equation as Eq. \eqref{eqa15}, but with a different parameter
\begin{equation}\label{eqc8aa}
 F_p=\frac{\tau}{2\eta d_s}\left(\frac{1}{1-2\pi  \chi _0}\right)^2\frac{\chi _0^2 a\Phi _0^2}{\lambda ^4(2\pi)^3},
\end{equation}
after introducing the same dimensionless units as before.

Hysteresis is developed when $F_p\ge 8$. For typical parameters for Nb superconductor $\xi\approx\lambda\approx 40$ nm, $\rho_n \approx 10^{-6}\ \rm{\Omega\cdot m}$ and $a=40$ nm at $B\approx 1$ T and $\chi_0=0.05$, $F_p> 8$ requires $\chi_0\tau>1$ ps. For the relaxation time of order $\chi_0\tau\approx 1\ \rm{\mu s}$, the effective viscosity is enhanced by a factor of $10^{6}$ compared to the bare BS one at $v< a/(\chi_0\tau)$. The effective critical current density for the whole system is given by
\begin{equation}\label{eqc9}
J_c\approx\left(\frac{1}{2-4\pi  \chi _0}\right)^2\frac{\chi _0c}{(2\pi )^4\lambda ^4}\frac{\Phi_0 a^2}{d_s+d_m}.  
\end{equation}
For $d_s=d_m=100$ nm , we obtain $J_c\approx 10^9\ \rm{A/m^2}$. The retrapping current $J_r$ is
\begin{equation}\label{eqc10}
J_r\approx \frac{1}{1-2\pi  \chi _0}\sqrt{\frac{\eta a d_s}{\pi \tau}}\frac{a c}{\lambda^2 4\pi^2} \frac{1}{d_s+d_m}.
\end{equation}
For the parameters used before and $\chi_0\tau=1\ \rm{{\mu s}}$, we estimate $J_r\approx 2\times 10^6\ \rm{A/m^2}$.

Let us discuss the optimal materials for the S and M layers. Superconductors with a smaller $\lambda$ are preferred because the critical current decreases as $\lambda^{-4}$. The smaller $\lambda$, the more nonuniform is the magnetic field distribution inside the M layers, hence stronger pinning. The viscosity in the branch with vortex polaron is proportional to $\tau$ while the critical current does not depend on $\tau$ for sufficiently large $\tau$. The slow magnetic dynamics can be realized in certain spin glasses, where the relaxation of magnetization is governed by a broad spectrum of time scale, with average time of the order $0.1\ \rm{\mu s}$ \cite{Uemura1985,Binder1986}. For $\rm{CuMn_{0.08}}$, $\chi_0\approx 0.002$ at $B=1$ T. \cite{Prejean1980} One may enhance $\chi_0$ by tuning the concentration of magnetic metal in alloys. \cite{Kouvel1961} One may also use superparamagnets with $\tau$ as large as $1$ s and with a huge $\chi_0$ due to large magnetic moments in superparamagnets \cite{Bean1959,Goldfarb1981,Cullity2008} and the recently synthesized cobalt-based and rare-earth-based single chain magnets with $\chi_0\approx 0.05$ at $B=1$ T and $10^{-6}\ \rm{s}<\chi_0\tau<10^{-4}\ \rm{s}$. \cite{Caneschi2001,Caneschi2002,Bogani2005,Bernot2006}

Now we discuss the optimal thickness of M and S layers. For $d_m\gg a$, we have $B_m(G>0)\approx \exp(-2\pi d_m/a)$ when $-d_m\ll z'\ll 0$ according to Eqs. (\ref{eqc3}) and (\ref{eqc4}). The magnetic induction and the magnetization are almost uniform in the lateral direction in the middle of the M layer. As a result, the pinning force becomes practically $d_m$ independent in this case. In other words, the pinning is effective only near the boundaries between S and M layers in the area of thickness of the order $a$. On the other hand, the Lorentz force is proportional $d_s$. Thus the effective critical current of the whole system $J_c$ is proportional to $1/(d_s+d_m)$ as described by Eq. (\ref{eqc9}). Therefore the thinner of both M and S layers, the higher is the critical current of the system.

The M/S multilayer structure is naturally realized in certain superconducting single crystals, such as (RE)Ba$_2$Cu$_3$O$_7$ \cite{Hor1987,Allenspach1995} and  $\rm{RuSr_2GdCu_2O_8}$ \cite{McLaughlin99}. For (RE)Ba$_2$Cu$_3$O$_7$, magnetic RE ions interact weakly with superconducting electrons because they are positioned between the superconducting layers. They order at very low N\'{e}el temperatures of the order $T_N\sim 1$ K. The polaronic mechanism is important at $T>T_N$, where spins are free. The London penetration depth of cuprate superconductors is large $\lambda\approx 200$ nm, thus the critical current is reduced significantly compared to that for Nb multilayer structure, because $J_c$ drops as $1/\lambda^4$. Another natural realization is the recently discovered iron-based superconductors, such as $\rm{(RE)FeAsO_{1-x}F_{x}}$, where $\rm{RE}$ ions ordered antiferromagnetically below $T_N\sim 1$K. \cite{Wang2008} In  $\rm{RuSr_2GdCu_2O_8}$ the magnetic moments order ferromagnetically above $T_c$ thus the dominant enhancement of vortex viscosity is due to the radiation of magnons \cite{Shekhter11,Shekhter11,szlin12a,szln12Spectrum}.

\section{Conclusions}
 In conclusions, vortices in magnetic superconductors polarize magnetic moments and become dressed and polaron-like. At low currents and long spin relaxation time the nonuniform polarization induced by vortices slows their motion at currents for which pinning by crystal lattice disorder becomes ineffective. As current increases above  the critical one,  vortices release nonuniform part of the polarization and the velocity as well as the voltage in the I-V characteristics jump to much higher values. At decreasing current vortices are retrapped by polarized magnetic moments at the retrapping current which is smaller than the critical one. The results of such polaronic mechanism are in qualitative agreement with the experimental data \cite{Yaron1996,Gammel2000} but measurements of the I-V characteristics are needed to establish the quantitative agreement and confirm validity of such a model for Er borocarbide. The polaronic mechanism should be at play also in Gd and Tb borocarbides superconductors in the commensurate SDW phase and strong effect  
may be present in Tm borocarbide above $T_N$.
 
We derive the response of the magnetic superconductors in the vortex state to the ac Lorentz force, $F_L(t)=F_{{\rm ac}}\sin(\omega t)$, taking into account the
polaronic effect. At low amplitudes of the driving force  $F_{{\rm ac}}$ the dissipation in the system is suppressed due to the enhancement of the effective viscosity at low frequencies and due to formation of the magnetic pinning at high frequencies $\omega$.  In the adiabatic limit with low frequencies $\omega$ and high amplitude of the driving force $F_{ac}$, the vortex and magnetic polarization form a vortex polaron when $F_L(t)$ is small. When $F_L$ increases, the vortex polaron accelerates and at a threshold driving force it dissociates, i.e.  the vortex motion and the magnetization relaxation decouple. When $F_L$ decreases, the vortex is retrapped by the background of remnant magnetization and they again form vortex polaron. This process repeats when $F_L(t)$ increases in the opposite direction. Remarkably, after dissociation, decoupled vortices move in the periodic potential induced by magnetization which remains for some  periods of time due to retardation of magnetization after the decoupling. At this stage vortices oscillate with high frequencies determined by the amplitude of the Lorentz force at the moment of dissociation.  

We propose to make multilayer system M/S where one can optimize superconducting and magnetic layers 
 to achieve  high critical current.
 
 \acknowledgments
The authors would like to thank P. Canfield, C. D. Batista, V. Kogan, V. Vinokur, D. Smith, A. Saxena, L. Civale and B. Maiorov for helpful discussions. This publication was made possible by funding from the Los Alamos Laboratory Directed Research and Development Program, project number 20110138ER.

%

\end{document}